\documentclass [12pt]{article}

\usepackage{tabularx}
\usepackage[english]{babel}
\usepackage[dvips]{graphicx}
\usepackage{indentfirst}
\usepackage{epsfig}
\begin{document}

\title{Quantum Equivalence of NC and YM Gauge Theories in $2$ D and Matrix Theory}

\author{Badis Ydri\footnote{Email: ydri@physik.hu-berlin.de.}~\footnote{ This work  is supported by a Marie Curie Fellowship from The Commision of the European Communities ( The Research Directorate-General ) under contract number MIF1-CT-2006-021797. The Humboldt-Universitat Zu Berlin preprint number is HU-EP-06/48.}}

\maketitle

\begin{abstract}
We construct noncommutative $U(1)$ gauge theory  on the fuzzy sphere $S^2_N$ as a unitary $2N\times 2N$ matrix model. In the quantum theory the model is 
equivalent to a  nonabelian $U(N)$ Yang-Mills theory on a $2$ dimensional lattice with $2$
 plaquettes. This equivalence  holds in the " fuzzy sphere" phase where we observe a $3$rd order phase transition between weak-coupling and strong-coupling  phases of the gauge theory. In the ``matrix'' phase we have a $U(N)$ gauge theory on a single point.
\end{abstract}

\section{Introduction}

A nonperturbative  regularization of noncommutative gauge theory in two dimensions is obtained by putting the theory on a fuzzy sphere $S^2_N$ \cite{thesis}. The lattice-like spacing parameter ( the inverse UV cut-off ) is proportional to $1/N$ where $N$ is the size of the matrix algebra. In this regulator the radius of the sphere provides an IR cut-off for the theory. The limit  $N{\longrightarrow}\infty$ is the continuum limit  \cite{madore}.

The differential calculus on the fuzzy sphere  is $3-$dimensional and as a consequence a spin $1$ vector field $\vec{A}$ is intrinsically $3-$dimensional. Each component $A_a$, $a=1,2,3$, is an element of $Mat_{N}$ with $N=n(L+1)$. Thus $U(n)$ symmetry will be implemented by $U(N)$ unitary transformations. On the fuzzy sphere $S^2_{L+1}$ it is not possible to split the vector field $\vec{A}$ in a gauge-covariant fashion into a tangent two-dimensional gauge field and a normal scalar fluctuation. We can only write a gauge-covariant expression for the normal field $\Phi$ as 
$\Phi=\frac{1}{2}(x_aA_a+A_ax_a+\frac{A_a^2}{\sqrt{c_2}})$ where $x_a$ are the coordinates on fuzzy $S^2_{L+1}$ defined by $x_a=L_a/\sqrt{c_2}$ with $c_2=\frac{L}{2}(\frac{L}{2}+1)$. $L_a$ are the generators of $SU(2)$ in the irreducible representation $\frac{L}{2}$. In the continuum limit $L{\longrightarrow}\infty $ the normal scalar field reduces to $\Phi =n_aA_a$. 

The $U(n)$ action on the fuzzy sphere $S^2_{L+1}$ reads (  with the identification $X_a=\alpha (L_a{\otimes}{\bf 1}_n+A_a)$ and with the normalization $Tr_N{\bf 1}_N=N$ )  

\begin{eqnarray}
S&=&N\bigg[-\frac{1}{4}Tr_N[X_a,X_b]^2+\frac{2i{\alpha}}{3}{\epsilon}_{abc}TrX_aX_bX_c\bigg]+\frac{Nm^2}{2c_2} Tr_N(X_a^2-{\alpha}^2c_2)^2.\label{main2}
\end{eqnarray}
In the continuum limit this action becomes ( with the gauge coupling constant defined by $g^2=\frac{1}{N^2{\alpha}^4}$  )
\begin{eqnarray}
S=\frac{1}{4g^2}\int \frac{d{\Omega}}{4\pi}tr_nF_{ab}^2-\frac{1}{4g^2}{\epsilon}_{abc}\int \frac{d{\Omega}}{4\pi}tr_n\big[F_{ab}A_c-\frac{i}{3}[A_a,A_b]A_c\bigg]+\frac{2m^2}{g^2}\int \frac{d{\Omega}}{4\pi}tr_n{\Phi}^2.\label{ck}
\end{eqnarray}
If we take the limit $m{\longrightarrow}\infty$ ( in other words we set $\Phi =0$ ) then the action will reduce to a $2-$dimensional pure $U(n)$ gauge theory
\begin{eqnarray}
S=\frac{1}{4g^2}\int \frac{d{\Omega}}{4\pi}tr_nF_{ab}^2=\frac{1}{4g^2}\int \frac{d{\Omega}}{4\pi}tr_n\bigg(i{\cal L}_aA_b-i{\cal L}_bA_a+{\epsilon}_{abc}A_c +i[A_a,A_b]\bigg)^2.\label{eq3}
\end{eqnarray}
In this last equation $A_a$ should  be understood as a $2-$dimensional gauge field ( in other words satisfying the constraint $n_aA_a=0$ ) and ${\cal L}_a=-i{\epsilon}_{abc}n_b{\partial}_c$.

The model (\ref{main2}) with $m=0$ was obtained in string theory limit in \cite{ars}. It was shown that it corresponds to a gauge theory on the fuzzy sphere in \cite{iso}. It was studied numerically in \cite{nishimura}. The model with $m=0$ and $m{\longrightarrow}\infty$ and
 for $U(1)$ groups was studied in one-loop perturbation theory in \cite{ref}. 

The basic prediction coming from  the Monte Carlo study \cite{ref11} is that noncommutative  $U(1)$ gauge theory in $2$ dimensions on the fuzzy sphere ( given by the above action ) behaves in the large $N$ limit like a commutative $U(N)$ in two dimensions on a lattice \cite{ref1}. This is true at least in the so-called "fuzzy sphere" phase of the model for large values of the gauge coupling constant. Indeed we observe in the simulation that this phase splits into two distinct regions corresponding to the weak and strong coupling phases of the gauge field which are separated by a third order phase transition. This transition seems to be consistent with that of a  one-plaquette  model \cite{gross}. It seems that deep inside the ``fuzzy sphere'' phase the model can still be understood as a $U(1)$ gauge theory on the sphere whereas in the ``matrix'' phase it is  a $U(N)$ gauge theory on a single point.

The aim of this article is to construct  models of noncommutative gauge theory on the fuzzy sphere which describe  this third order phase transition although in general these models will coincide with (\ref{main2}) only in the continuum limit. An alternative  approach to gauge theory on the fuzzy sphere is given in \cite{steinackers2}. 
\section{The Model}

The main idea is to reparametrize  the gauge field on ${S}^2_N$ in terms of a single matrix $W$ which contains all the tangent degrees of freedom. This  $W$ we call  the " fuzzy link variable" . Thus we will have the coordinate transformation $
(A_1,A_2,A_3){\longrightarrow}(W,\Phi)$. Let us
 introduce the $2N{\times}2N$ idempotent 
\begin{eqnarray}
{\gamma}=\frac{1}{N}({ 1}+2{\sigma}_aL_a)~,~{\gamma}^2=1
\end{eqnarray}
 where ${\sigma}_a$ are the usual Pauli matrices. It has eigenvalues $+1$ and $-1$ with multiplicities $N+1$ and $N-1$ respectively. We introduce the covariant derivative $D_a=L_a+A_a$ through a gauged idempotent ${\gamma}_D$ as follows 
\begin{eqnarray}
&&{\gamma}_D=\hat{\gamma}\frac{1}{\sqrt{\hat{\gamma}^2}}~,~\hat{\gamma}=\frac{1}{N}(1+2{\sigma}_aD_a)~,~~\hat{\gamma}^2=1+\frac{8\sqrt{c_2}}{N^2}\Phi+\frac{2}{N^2}{\epsilon}_{abc}{\sigma}_cF_{ab}.\label{eq0}
\end{eqnarray}
In above $F_{ab}=i[D_a,D_b]+{\epsilon}_{abc}D_c=i[L_a,A_b]-i[L_b,A_a]+{\epsilon}_{abc}A_c+i[A_a,A_b]$. 
Clearly ${\gamma}_D$ has the same spectrum as ${\gamma}$. Thus there exists a $U(2N)$ unitary transformation $U$ such that ${\gamma}_D=U\gamma U^+$. The  crucial observation is that if we let $T{\in}U(N+1){\times}U(N-1)$ then $U T\gamma T^+ U^+=U\gamma U^+$ since $T\gamma T^+=\gamma $. In other words  $U$ is an element of the $d_N-$Grassmannian manifold $G_N=U(2N)/(U(N+1){\times}U(N-1))$ and hence it ( or equivalently ${\gamma}_D$ )  contains the correct number of degrees of freedom  $
d_N$ which is found in a gauge theory on the fuzzy sphere without normal scalar field. Indeed  $
d_N=4N^2-(N+1)^2-(N-1)^2=2N^2-2$.  Thus ${\gamma}_D$ can be identified with the equivalence class $[U]=\{UT,T{\in}U(N+1)\times U(N-1)\}$.

This fact can also be seen by writing down the measure
 in terms of $U$ explicitly. A direct calculation gives  
$Tr_{2N}(d{\gamma}_D)^+(d{\gamma}_D)
=8\sum_{i=1}^{N+1}\sum_{j=N+2}^{2N}(dU^+U)_{ij}(U^+dU)_{ji}$. 
It is obvious that only $d_N=2N^2-2$ degrees of freedom of $U$ are involved. In the large $N$ limit we can see from (\ref{eq0}) that ${\gamma}_D{\longrightarrow}\gamma$ and hence $U{\longrightarrow}{\bf 1}_{2N}$. Thus in this limit we can write $U=1+i\frac{\Lambda}{N}+O(\frac{{\Lambda}^2}{N^2})$ and as a consequence 
\begin{eqnarray}
Tr_{2N}(d{\gamma}_D)^+(d{\gamma}_D)
=\frac{8}{N^2}Tr_{N-1}(d{\Lambda}_{12})^+d{\Lambda}_{12}+O(\frac{{\Lambda}^3}{N^3}).
\end{eqnarray}
In above   ${\Lambda}_{12}$ is  the off-diagonal upper block of ${\Lambda}$ and ${\Lambda}_{12}^+$ is  the off-diagonal lower block of ${\Lambda}$.  ${\Lambda}_{12}$ is an $(N+1)\times (N-1)$ matrix while ${\Lambda}_{12}^+$ is an $(N-1)\times (N+1)$ matrix. We have then in the limit the measure
\begin{eqnarray}
\int_{G_N} d{\gamma}_D ~{\equiv} \int_{G_N} d[U]{\propto}\int d{\Lambda}_{12}d{\Lambda}_{12}^+.\label{measrr}
\end{eqnarray}
Integration over $G_N$ means that we integrate over all idempotents ${\gamma}_D$ which have $N+1$ eigenvalues equal $+1$ and $N-1$ eigenvalues equal $-1$. A general idempotent with this property can be parametrized in terms of a  unitary matrix $U$ as $U\gamma U^+$. Thus  ( in the basis in which $\gamma$ is diagonal with ${\bf 1}_{N+1}$ in the first block and $-{\bf 1}_{N-1}$ in the second block )  we integrate only over the $d_N=2N^2-2$ degrees of freedom of the unitary matrix $U$ which correspond to ${\gamma}_D$. 
As we will see only these degrees of freedom will effectively appear in the action. 

A general idempotent with the above property can also be parametrized in terms of $3$ hermitian matrices $D_a$ as in equation (\ref{eq0}). Indeed we can check that we have ( with ${\gamma}_D$ given in terms of $D$'s by equation (\ref{eq0}) ) 
\begin{eqnarray}
Tr_{2N}(d{\gamma}_D)^+d{\gamma}_D=\frac{8}{N^2}Tr_N(dD_a)^+dD_a-\frac{8}{N^2}Tr_N(d\Phi)^+d{\Phi}+O(\frac{1}{N^3}).
\end{eqnarray}
Then in the large $N{\longrightarrow}\infty $ limit it is obvious that   we have the  measure

\begin{eqnarray}
\int dA_1dA_2dA_3 \equiv \int dD_1dD_2dD_3~{\propto}~\int_{G_N} d{\gamma}_D\int d\Phi. \label{meas}
\end{eqnarray}
In the limit $m{\longrightarrow}\infty $ we have $\Phi =0$ and hence we obtain  instead 
\begin{eqnarray}
 \int dD_1dD_2dD_3{\delta}(\Phi)~{\propto}~\int_{G_N} d{\gamma}_D.\label{meas1}
\end{eqnarray}
Thus for consistency  we must set $\Phi=0$ in $\hat{\gamma}^2$. We can then compute ( with the covariant coordinates $x_a^D$ defined by $x_a^D=\frac{2}{N}D_a$ and $\epsilon \sigma F\equiv{\epsilon}_{abc}{\sigma}_cF_{ab}$ ) the following expansion
\begin{eqnarray}
{\gamma}_D =\frac{1}{N}+{\sigma}x^D-\frac{1}{N^2}({\sigma}x^D)({\epsilon}{\sigma}F)-\frac{1}{N^3}({\epsilon}{\sigma}F)+\frac{3}{2N^4}({\sigma}x^D)({\epsilon}{\sigma}F)^2+\frac{3}{2N^5}({\epsilon}{\sigma}F)^2+O(\frac{1}{N^6}).
\end{eqnarray}
Let $U$ and  $V$ be  $2N\times 2N$ unitary matrices which are in the Grassmannian manifold $G_N$ and let us consider the following   path integral 
\begin{eqnarray}
Z[\gamma ,{\lambda}]=\int_{G_N} d[U] \int_{G_N} d[V] ~\exp(-S_P)~,~S_P=-\frac{4N^2}{\lambda}+\frac{N}{\lambda}Tr_{2N}U\gamma U^+ V\gamma V^+ +{\rm h.c}.\label{path}
\end{eqnarray}
As we have explained integration over $G_N$ means that we integrate only over the $2d_N=2(2N^2-2)$  degrees of freedom of the unitary matrices $U$ and $V$ which will effectively appear in the action. 

The matrix $U\gamma U^+$ can always be parametrized as $U\gamma U^+={\gamma}_D$ where ${\gamma}_D$ is the idempotent given by (\ref{eq0}). For small $U$ near the identity given by $U=1+i\frac{\Lambda}{N}+... $  we can show that ${\gamma}_D$ ( and hence the action )
will only depend on the off-diagonal blocks ${\Lambda}_{12}$ and ${\Lambda}_{12}^+$ of $\Lambda$. We will only integrate over these components in the measure. 
Similarly  the matrix $V\gamma V^+$ can  be parametrized as $V\gamma V^+={\gamma}_B$ where ${\gamma}_B$ is another idempotent given by equation (\ref{eq0}) but in terms of a different covariant derivative $B_a=L_a+A_a^{'}$ with a new gauge field $A_a^{'}$ and normal component ${\Phi}^{'}$.  As before  we write $V=1+i\frac{{\Lambda}^{'}}{N}+...$ in the large $N$ limit and again the action  will only depend on the off-diagonal blocks ${\Lambda}^{'}_{12}$ and $({\Lambda}^{'}_{12})^+$ of ${\Lambda}^{'}$. We will only integrate over these components in the measure.

The action $S_P$ is invariant under all $U(2N)$ unitary transformations of the form $U{\longrightarrow}gU$, $V{\longrightarrow}gV$ which is only possible beacuse of the doubling of gauge fields. The idempotents ${\gamma}_D$ and ${\gamma}_B$ will therefore transform  as ${\gamma}_D{\longrightarrow}g{\gamma}_Dg^+$ and ${\gamma}_B{\longrightarrow}g{\gamma}_Bg^+$ respectively. Thus the measures $d{\gamma}_D$ and $d{\gamma}_B$ are also $U(2N)-$symmetric and therefore the full path integral (\ref{path}) is also $U(2N)-$symmetric.

Next we compute the action $S_P$ in the configuration $W={\gamma}_D {\gamma}_B$. 
Let $G_{ab}$ and $x_a^B$ be the curvature and the covariant coordinates associated with the covariant derivative $B_a$. We obtain after a short calculation 
\begin{eqnarray}
S_P&=&\frac{1}{N\lambda}Tr_N\bigg[-32c_2
+2N^2(x_a^D+x_a^B)^2
+\frac{8i}{N^2}[D_a-B_a,D_b-B_b](F_{ab}+G_{ab})
-\frac{8}{N^2}(F_{ab}+G_{ab})^2\nonumber\\
&+&
\frac{6}{N^2}\{x_a^D,x_a^B\}(F_{cd}^2+G_{cd}^2)-\frac{4}{N^2}x_a^DF_{cd}x_a^BG_{cd}-\frac{4}{N^2}x_a^BF_{cd}x_a^DG_{cd}\nonumber\\
&+&\frac{2}{N^2}x_a^D(\epsilon F)_ax_b^B(\epsilon G)_b+\frac{2}{N^2}x_a^D(\epsilon F)_bx_b^B(\epsilon G)_a+\frac{2}{N^2}x_a^B(\epsilon F)_ax_b^D(\epsilon G)_b+\frac{2}{N^2}x_a^B(\epsilon F)_bx_b^D(\epsilon G)_a\nonumber\\
&-&\frac{3}{N^2}\bigg([x_a^D,x_b^B]+[x_a^B,x_b^D]\bigg)\bigg((\epsilon F)_a(\epsilon F)_b+(\epsilon G)_a(\epsilon G)_b\bigg)
+O(\frac{1}{N^3})\bigg].\label{path1}
\end{eqnarray}

These are the first few terms of the $U(1)$ action in this fuzzy plaquette model. The plaquette variable is the unitary $2N\times 2N$ matrix given by $W={\gamma}_D{\gamma}_B$. This fuzzy plaquette $U(1)$ model is more complicated than the original fuzzy $U(1)$ model (\ref{main2}) although the two actions  (\ref{main2}) and (\ref{path1}) have the same continuum limit as we will now show.

We show this result in  two steps. First we  compute the continuum limit of the above classical action then we integrate out one of the gauge fields. We will see in particular that the effect of the first non-trivial term $2N^2Tr_N(x_a^D+x_a^B)^2$ is such that the path integration  over $B_a$ is dominated by the configuration   $B_a=-D_a$. 
Indeed in the continuum large $N$ limit in which we keep $N^2\lambda$ fixed we can use in the classical action $S_P$ the limits $x_a^D,x_a^B{\longrightarrow}n_a$ where $n_a$ are the global coordinates on the sphere  and hence $S_P$ reduces to 
\begin{eqnarray}
S_P&=&-\frac{32c_2}{\lambda}+\frac{2N^2}{\lambda}\int \frac{d\Omega}{4\pi}
(x_a^D+x_a^B)_*^2+\frac{4}{N^2\lambda}\int \frac{d\Omega}{4\pi}\bigg[F_{ab}^2+G_{ab}^2-6F_{ab}G_{ab}+2({\epsilon}_{abc}n_c  F_{ab})({\epsilon}_{abc}n_c G_{ab})\bigg]\nonumber\\
&=&-\frac{32c_2}{\lambda}+\frac{2N^2}{\lambda}\int \frac{d\Omega}{4\pi}
(x_a^D+x_a^B)^2_*
+\frac{4}{N^2\lambda}\int \frac{d\Omega}{4\pi}(F_{ab}-G_{ab})^2
.\label{path11}
\end{eqnarray}
In above we have used the fact that on the sphere $({\epsilon}_{abc}n_c  F_{ab})({\epsilon}_{abc}n_cG_{ab})=2F_{ab}G_{ab}$ and $*$ stands for the star product on the fuzzy sphere which still appear  in the second term. The path integral over the $B_a$ is clearly seen to be dominated by the configuration $B_a=-D_a$\footnote{This means in particular that the unitary matrix $W={\gamma}_D{\gamma}_B$ approaches in the quantum theory $-{\bf 1}_{2N}$ in contrast with the classical limit $W{\longrightarrow}{\bf 1}_{2N}$.}. Therefore  we obtain the action ( modulo a constant term )

\begin{eqnarray}
S_P=\frac{N}{\lambda}Tr_{2N}W+{\rm h.c}=\frac{16}{N^2\lambda}\int \frac{d\Omega}{4\pi} F_{ab}^2.\label{F}
\end{eqnarray}
This action is essentially the $U(1)$ action obtained from (\ref{main2}) ( see equation (\ref{eq3}) ) provided we make the following identification 
\begin{eqnarray}
\frac{16}{ N^2{\lambda}}\equiv\frac{1}{4g^2}\equiv \frac{\tilde{\alpha}^4}{4}\equiv \frac{\bar{\alpha}^4}{4N^2}.\label{mmm}
\end{eqnarray}
Hence the fuzzy  $U(1)$ action (\ref{main2}) with fixed coupling constant $\alpha$ corresponds in this particular limit to the fuzzy $U(1)$ plaquette  action (\ref{path1}) in the weak regime ${\lambda}{\longrightarrow}0$ and agreement between the two is expected only for weak couplings ( small values of $\lambda$ or equivalently large values of $\bar{\alpha}$ ).  This is what we observe numerically \cite{ref11}.

\section{The solution}
The path integral (\ref{path}) is invariant under all $U(2N)$ unitary transformations of the form $U{\longrightarrow}gU$, $V{\longrightarrow}gV$ or equivalently ${\gamma}_D{\longrightarrow}g{\gamma}_Dg^+$, ${\gamma}_B{\longrightarrow}g{\gamma}_Bg^+$. This symmetry can be fixed as follows. We perform the following $U(2N)$ transformation $U{\longrightarrow}U^{'}=gU$ with $g=V^+$. For fixed $V$ we obtain the measure $\int _{G_N} dU=\int_{G_N} dU^{'}$ and the action $S_P$ will only depend on $U^{'}$. The integral over $V$ can be done and one ends up with the path integral ( by dropping also the primes )
\begin{eqnarray}
Z[\gamma ,{\lambda}]&=&\int_{G_N} d[U] ~\exp\bigg(\frac{4N^2}{\lambda}-\frac{N}{\lambda}Tr_{2N}\big(\gamma U\gamma U^+   +{\rm h.c}\big)\bigg)\nonumber\\
&{\equiv}&\int_{G_N} d{\gamma}_D ~\exp\bigg(\frac{4N^2}{\lambda}-\frac{N}{\lambda}Tr_{2N}\big(\gamma {\gamma}_D   +{\rm h.c}\big)\bigg).\label{gammad}
\end{eqnarray}
This is still invariant under  $U(N+1)\times U(N-1)$ unitary transformations  $U{\longrightarrow}gU$. This  is the gauge symmetry we want on the fuzzy sphere with $N\times N$ matrices. Let us recall that we want a $U(N)$ gauge symmtery on the fuzzy sphere but in this construction where we are using $2N\times 2N$ matrices it is only natural to obtain a gauge group which  is twice as large. Remark also that we do not  get precisely $U(N)$ but the groups $U(N+1)$ and $U(N-1)$. In the large $N$ limit this becomes unimportant.

Apart from the restricted integration which is performed on the Grassmannian manifold $G_N$ the  path integral (\ref{gammad}) looks very much like a $U(2N)$ one-plaquette model. Indeed the  matrix $W=\gamma {\gamma}_D$ is a unitray $2N\times 2N$ matrix. Let us then write (\ref{gammad}) in the equivalent form
\begin{eqnarray}
Z[\gamma ,{\lambda}]&=&\int dW \int_{G_N} d{\gamma}_D {\delta}(W-\gamma {\gamma}_D) ~\exp\bigg(\frac{4N^2}{\lambda}-\frac{N}{\lambda}Tr_{2N}\big(W   +{\rm h.c}\big)\bigg)\nonumber\\
&=&\int dW ~I(W) ~\exp\bigg(\frac{4N^2}{\lambda}-\frac{N}{\lambda}Tr_{2N}\big(W   +{\rm h.c}\big)\bigg).\label{gammad2}
\end{eqnarray}
We can therefore   approximate the partition function of noncommutative $U(1)$ gauge theory on the fuzzy sphere given by (\ref{path}) with the following path integral
\begin{eqnarray}
&&Z[\gamma ,{\lambda},M^2]
=\int dW ~I(W,M^2) ~\exp\bigg(\frac{4N^2}{\lambda}-\frac{N}{\lambda}Tr_{2N}\big(W   +{\rm h.c}\big)\bigg)\nonumber\\
&&I(W,M^2)=\int_{G_N} d{\gamma}_D\exp \bigg(-M^2Tr_{2N}\big(W-\gamma {\gamma}_D\big)^+\big(W-\gamma {\gamma}_D\big)\bigg).\label{gammad3}
\end{eqnarray}
Instead of the delta function in (\ref{gammad2}) which implements the constraint $W-\gamma {\gamma}_D=0$ we  add a new term to the action with large positive coupling constant $M^2$ which implements this constraint only approximately. As it turns out the  integral over $G_N$ in (\ref{gammad3}) is much easier to compute ( at least in large $N$ ) than the original corresponding 
path 
integral $I(W)$  in (\ref{gammad2}). 

We remark that by dropping the requirement that $M^2$ must be large 
we get a model with an enlarged phase space. Indeed the space of parameters of the theory becomes the two dimensional quadrant $\lambda {\geq}0$, $M^2{\geq}0$ instead of the original positive real line. 
An alternative way of thinking about (\ref{gammad3}) is as follows. 
The  one-parameter family of models $Z_M\equiv Z[\gamma,{\lambda},M^2]$ interpolate between the model $Z_0$ which corresponds to a true $U(2N)$ one-plaquette model and $Z_{\infty}$ which is the path integral of the noncommutative $U(1)$ gauge theory on the fuzzy sphere given by (\ref{gammad2}) or equivalently (\ref{path}). 
In other words holding $N$ fixed and taking the limit $M^2{\longrightarrow}\infty$ reproduces the original path integral (\ref{gammad2}). 

It is hopped that the models $Z_M$  with intermediate large values of $M^2$ will capture some of the most important features of noncommutative $U(1)$ gauge theory on the fuzzy sphere ( at least in the "fuzzy sphere" phase ) seen in the Monte Carlo study of the action  (\ref{main2}). 
In particular we will consider in the following  the double scaling limits $M^2,N{\longrightarrow}\infty$ keeping $M^2/N$ fixed. 
By construction the models $Z_M$ coincide in these double scaling limits with the limit $N{\longrightarrow}\infty$ of (\ref{gammad2}). So these models have the correct continuum limit and hence they are also path integrals of noncommutative $U(1)$ gauge theories on the fuzzy sphere. However quantum mechanically these models for finite fixed values of the ratio $M^2/N$ are found to behave differently from   (\ref{gammad2})  which corresponds to the ratio  $M^2/N{\longrightarrow}\infty$. As we will see the limit $M^2/N{\longrightarrow}0$ is also different.

The matrix $\gamma {\gamma}_D$ is a unitray matrix which in the continuum large $N$ limit tends ( by equation (\ref{eq0}) ) to the identity matrix ${\bf 1}_{2N}$. Indeed in this limit both ${\gamma}$ and ${\gamma}_{D}$ approach the usual chirality operator $\gamma =n_a{\sigma}_a$ and hence $\gamma {\gamma}_D{\longrightarrow}{\bf 1}_{2N}$. Equivalently the unitary matrix $U$ will approach in the continuum limit the identity ${\bf 1}_{2N}$ as $U=1+i\frac{\Lambda}{N}+O(\frac{{\Lambda}^2}{N^2})$. In this limit it is also seen   by equation (\ref{measrr})   that  the measure $\int_{G_N} d{\gamma}_D\equiv \int_{G_N} d[U]$ depends  only on the off-diagonal blocks ${\Lambda}_{12}$ and ${\Lambda}_{21}$ with ${\Lambda}_{21}^{+}={\Lambda}_{12}$. Let $W_1$ and $W_2$ be the diagonal blocks of $W$ in the basis where $\gamma$ is diagonal. We denote by $W_{12}$ and $W_{21}$  the off-diagonal blocks.  Thus
in the large $N$ limit the second line of (\ref{gammad3}) takes the form

\begin{eqnarray}
I(W,M^2)&=&\int_{G_N} d{\gamma}_De^{-M^2Tr_{2N}\big(2-W^+\gamma {\gamma}_D-W{\gamma}_D\gamma \big)}\nonumber\\
&{\propto}&\int d{\Lambda}_{12}d{\Lambda}_{21}e^{-M^2Tr_{2N}\big(W-1\big)^+\big(W-1\big)-\frac{2iM^2}{N}Tr_{N+1}\big(W_{21}^+-W_{12}\big){\Lambda}_{21}-\frac{2iM^2}{N}Tr_{N-1}\big(W_{12}^+-W_{21}){\Lambda}_{12}}.\label{w1w2-0}\nonumber\\
\end{eqnarray}
There are two cases to consider. 

{\underline{Case I :}} This corresponds to the special case  $M^2=0$. By the requirement of continuity this point should be reached in the double scaling limits $M^2,N{\longrightarrow}\infty$ keeping $\frac{M^2}{N}$ fixed such that $\frac{M^2}{N}{\longrightarrow}0$. The above integral (\ref{w1w2-0}) becomes  
\begin{eqnarray}
I(W,M^2)&=&e^{-M^2Tr_{2N}\big(W-1\big)^+\big(W-1\big)}.
\end{eqnarray}
Thus we end up with the partition function
\begin{eqnarray}
&&Z[\gamma ,{\lambda},M^2]{\propto}
\int dW ~~\exp\bigg(\frac{4N^2}{{\lambda}_M}-\frac{N}{{\lambda}_M}Tr_{2N}\big(W   +{\rm h.c}\big)\bigg)~,~\frac{1}{{\lambda}_M}=\frac{1}{\lambda}-\frac{M^2}{N}.\label{gammad4-0}
\end{eqnarray}

{\underline{Case II :}} This is the generic case of the double scaling limits $M^2,N{\longrightarrow}\infty$ keeping $\frac{M^2}{N}$ finite and fixed. Extrapolation of the results to the case $\frac{M^2}{N}{\longrightarrow}\infty$ will by definition correspond    to the model (\ref{gammad2}). However extrapolation  of the results to  $\frac{M^2}{N}{\longrightarrow}0$ will not reproduce case I as we will see. In these double scaling limits the integral (\ref{w1w2-0}) becomes
\begin{eqnarray}
I(W,M^2)
&{\propto}&e^{-M^2Tr_{N+1}\big(2-W_1-W_1^+\big)-M^2Tr_{N-1}\big(2-W_2-W_2^+\big)}{\delta}\big(W_{21}^+-W_{12}\big){\delta}\big(W_{12}^+-W_{21}).\label{w1w2}
\end{eqnarray}
In above ${W}_1$ is an $(N+1)\times (N+1)$ matrix, ${W}_2$ is an $(N-1)\times (N-1)$ matrix and ${W}_{12}$, ${\Lambda}_{12}$ are $(N+1)\times (N-1)$ matrices whereas ${W}_{21}$, ${\Lambda}_{21}$ are $(N-1)\times (N+1)$ matrices. 

Since in these limits $M^2$ is large proportional to $N$  we also see that we have   $W{\longrightarrow}{\bf 1}_{2N}$ or equivalently $W_1{\longrightarrow}{\bf 1}_{N+1}$ and $W_2{\longrightarrow}{\bf 1}_{N-1}$ and hence  $W_{12}+W_{21}^+{\longrightarrow}0$ which follows from the identity $W_1^+W_{12}+W_{21}^+W_2=0$. Together with the delta function  ${\delta}\big(W_{21}^+-W_{12}\big){\delta}\big(W_{12}^+-W_{21})$ appearing in the last line of equation (\ref{w1w2}) we can conclude that  the off-diagonal blocks   $W_{12}$ and $W_{21}$ of $W$  will approach  zero in this double scaling limit. By neglecting also the edge effects in the large $N$ limit we can take ${W}_{1}$ and ${W}_{2}$ to be $N\times N$  matrices. The partition function given by the first line of (\ref{gammad3}) becomes
\begin{eqnarray}
&&Z[\gamma ,{\lambda},M^2]{\propto}[Z({\lambda}_M)]^2~,~Z({\lambda}_M)
=\int dW_1 ~~\exp\bigg(\frac{2N^2}{{\lambda}_M}-\frac{N}{{\lambda}_M}Tr_{N}\big(W_1   +{\rm h.c}\big)\bigg).
\label{gammad4}
\end{eqnarray}
Therefore in these double scaling  continuum limits in which $M^2,N{\longrightarrow}\infty$ keeping $M^2/N$  fixed we can  set  $I(W)$ in the first line of equation (\ref{gammad3}) equal to $1$ and  replace $W$ with the  matrix  obtained  by taking the diagonal parts $W_1$ and $W_2$  to be two arbitrary independent unitary $N\times N$ matrices  while allowing the off-diagonal parts  $W_{12}$ and $W_{21}$ to go to zero.  In this approximation   we can drop the restriction that these unitary matrices $W_1$ and $W_2$ must be close to ${\bf 1}_{N}$ since we have replaced the coupling constant $\lambda$ with ${\lambda}_M$.

The path integral of  a $2-$dimensional $U(N)$ gauge theory in the axial gauge $A_1=0$  on a lattice with volume $V$ and lattice spacing $a$ is given by $Z({\lambda}_M )^{V/a^2}$.  
Therefore we can see that  the partition function $Z(\gamma,\lambda,M^2)$ of  noncommutative $U(1)$ gauge theory on the fuzzy sphere  is proportional to the partition function of   a $2-$dimensional $U(N)$ gauge theory in the axial gauge  $A_1=0$  on a lattice with two plaquettes. This is true at least in the weak coupling region of the phase space. Indeed since ${\lambda}_M {\geq}0$ we must have 
\begin{eqnarray}
{\lambda}{\leq}\frac{N}{M^2}.\label{range5}
\end{eqnarray}

\paragraph{The matrix phase}
The fact that the one-plaquette model (\ref{gammad4}) can be defined for all values of the coupling constant ${\lambda}_M$ in the range $[0,\infty [$ leads to the above restriction on the allowed values of the gauge coupling constant $\lambda$. This can be understood as follows. The partition function $Z(\gamma,\lambda,M^2)$ corresponds to noncommutative $U(1)$ gauge theory on the fuzzy sphere only for values of the coupling constant $\lambda$ below the  upper value $\frac{N}{M^2}$ where we can make sense of this model as a one-plaquette model. For  ${\lambda}{>}\frac{N}{M^2}$ the model $Z(\gamma,\lambda,M^2)$ is presumably in the so-called "matrix" phase where there is no an underlying stable sphere as a spacetime and the gauge theory is defined on a single point.  

This picture is consistent with our previous results \cite{ref11,ref}. Indeed we found in the one-loop calculation as well as in numerical simulation that the model (\ref{main2}) undergoes a first order phase transition from the "fuzzy sphere" phase to a "matrix" phase where the fuzzy sphere vacuum collapses under quantum fluctuation. The fuzzy sphere phase is defined only for values of the coupling constant $\lambda$ such that  ( $m^2$ is the mass parameter appearing in (\ref{main2}) )
\begin{eqnarray}
\lambda{\leq}{\lambda}_*^{S^2{\longrightarrow}\{0\}}=\frac{8}{N^2}(m^2+\sqrt{2}-1). \label{pred}
\end{eqnarray}

\paragraph{The one-plaquette $3$rd order phase transition} In this section we solve  the models given by (\ref{gammad4-0}) and (\ref{gammad4}) in the large $N$ limit. 
The only difference between the two models lies in the fact that (\ref{gammad4-0}) is a $U(2N)$ plaquette model while (\ref{gammad4}) is a $U(N)$ plaquette model and hence we will concentrate on the detail of (\ref{gammad4}) which describes   noncommutative $U(1)$ gauge theory on the fuzzy sphere. 

We can immediately diagonalize the matrix $W_1$ in (\ref{gammad4}) by writing $W_1=TDT^{+}$ where $T$ is some $U(N)$ matrix and $D$ is diagonal with elements equal to the eigenvalues ${\theta}_i+\pi$ of $W_1$\footnote{The addition of the angle $\pi$  reflects the fact that the unitary matrix $W$ approaches in the quantum theory $-{\bf 1}_{2N}$ and not ${\bf 1}_{2N}$ for very weak couplings. See the footnote on page $5$.}. In other words $D_{ij}={\delta}_{ij}({\theta}_i+\pi)$. The integration over $T$ can be done trivially and one ends up with the path integral
\begin{eqnarray}
Z({\lambda}_M)=\int {\prod}_{i=1}^Nd{\theta}_i ~e^{NS_N}~,~S_N=\frac{2N}{{\lambda}_M}+ \frac{2}{{\lambda}_M}\sum_{i=1}^N \cos {\theta}_i+\frac{1}{2N}\sum_{i{\neq}j}\ln \bigg(\rm sin\frac{{\theta}_i-{\theta}_j}{2}\bigg)^2.\label{1p}
\end{eqnarray}
The last term is due to the Vandermonde determinant. In the large $N$ limit we can resort to the method of steepest descent to evaluate the path integral $Z(\lambda)$ . The partition function will be dominated by the solution of the equation $\frac{dS_N}{d{\theta}_i}=0$ which is a minimum of the action $S_N$. In the large $N$ limit we  introduce a density of eigenvalues ${\rho}(\theta)$ which is positive definite and normalized to one.  Thus sums will be replaced by $\sum_{i}=N\int d\theta {\rho}(\theta)$. The saddle point solution must satisfy the equation of motion
\begin{eqnarray}
\frac{2}{{\lambda}_M}\sin {\theta}=\int_{}^{}d{\tau} {\rho}({\tau})\cot\frac{{\theta}-{\tau}}{2}.
\end{eqnarray}
By using the expansion $\cot\frac{{\theta}-{\tau}}{2}=2\sum_{n=1}^{\infty}\big(\sin n\theta \cos n\tau-\cos n\theta \sin n\tau \big)$ we can solve this equation quite easily in the strong-coupling phase and one finds the solution \cite{gross}
\begin{eqnarray}
{\rho}({\theta})=\frac{1}{2{\pi}}+\frac{1}{\pi  {\lambda}_M }\cos  {\theta}.\label{strong}
\end{eqnarray}
However it is obvious that this solution makes sense only  where the density of eigenvalues is positive definite. This density of eigenvalues ${\rho}$ is positive definite for all values of ${\theta}$ iff  ${\lambda}_M{>}2$. The critical value of the coupling constant ${\lambda}_M$ is then seen to occur at ${\lambda}_{M*}=2$. In terms of the coupling constant $\lambda$ we obtain the critical value
\begin{eqnarray}
{\lambda}_{*}^u=\frac{2}{1+2\frac{M^2}{N}}.\label{u}
\end{eqnarray}
 From (\ref{range5}) we then see that for the weak coupling phase we have ${\lambda}{\leq}{\lambda}_{*}^u{\leq}\frac{N}{M^2}$. In the strong coupling phase ${\lambda}_{*}^u{\leq}{\lambda}{\leq}\frac{N}{M^2}$. 

In the weak regime we obtain the solution \cite{gross}

\begin{eqnarray}
{\rho}({\theta}_{})=\frac{2}{{\pi}{\lambda}_M^{}}\cos \frac{{\theta}_{}}{2}\sqrt{\frac{{\lambda}_M^{}}{2}-\sin^2\frac{{\theta}_{}}{2}}~,~
\sin\frac{{\theta}_{*}}{2}=\sqrt{\frac{{\lambda}_M^{}}{2}}.\label{ik}
\end{eqnarray}
Again the critical value of the coupling constant ${\lambda}_M$ is seen to occur at ${\lambda}_{M*}=2$ when the critical angle ${\theta}_*$ approaches $\pi$. At this point the eigenvalues $\theta$ of $W_1$ will fill the whole unit circle.

For very strong couplings the density of eigenvalues (\ref{strong}) becomes a uniform distribution while in the very weak coupling region ( which corresponds to very small angles $\theta$ ) the density of eigenvalues (\ref{ik}) reduces to Wigner semicircle law, viz 
\begin{eqnarray}
&&{\rho}(\theta)=\frac{1}{\pi{\lambda}_M}\sqrt{2{\lambda}_M-{\theta}^2}.\label{rho}
\end{eqnarray}

We can use this last equation  to compute  the free energy and specific heat for small values of the coupling constant $\lambda$ with excellent agreement with the simulation results of (\ref{main2}). The free energy is given by
\begin{eqnarray}
\frac{F_w}{N^2}=2\times\frac{S_N}{N}&=&\frac{8}{{\lambda}_M}-\frac{2}{{\lambda}_M}\int_{-{\theta}_*}^{{\theta}_*} d\theta {\rho}(\theta){\theta}^2+\int_{-{\theta}_*}^{{\theta}_*}  d\theta {\rho}(\theta)\int_{-{\theta}_*}^{{\theta}_*}  d\alpha {\rho}(\alpha)\ln\bigg(\frac{\theta -\alpha}{2}\bigg)^2\nonumber\\
&=&\frac{8}{{\lambda}_M}-1+\ln\frac{{\lambda}_M}{2}+c_1.
\end{eqnarray}
The factor of $2$ which multiplies $\frac{S_N}{N}$ comes from the fact that we have two identical independent one-plaquette models contributing to $F$  in accordance with (\ref{gammad4}). In order to compute the specific heat  we implement the scaling transformations $F{\longrightarrow}\frac{F}{T}$ and ${\lambda}_M{\longrightarrow}T{\lambda}_M$. The specific heat is then defined by
\begin{eqnarray}
C_{\rm v}=\bigg(T^3\frac{{\partial}^2F}{{\partial}T^2}\bigg)_{T=1}.
\end{eqnarray}
A straightforward calculation yields the very simple result
\begin{eqnarray}
\frac{C_{\rm v}}{N^2}=1.
\end{eqnarray}
This is precisely what we see numerically in the fuzzy sphere phase. The value $1$ emerges from the fact that we have two plaquettes. As it turns out this result is valid  throughout the weak-coupling phase, i.e for all ${\lambda}_M{\leq}2$.

In the regime of strong couplings the free energy and specific heat are computed using the distribution of eigenvalues (\ref{strong}). We find

 \begin{eqnarray}
\frac{F_s}{N^2}=2\times\frac{S_N}{N}&=&\frac{4}{{\lambda}_M}+\frac{4}{{\lambda}_M}\int_{-\pi}^{\pi} d\theta {\rho}(\theta)\cos {\theta}+\int_{-\pi}^{\pi}  d\theta {\rho}(\theta)\int_{-\pi}^{\pi}  d\alpha {\rho}(\alpha)\ln\bigg(\sin \frac{\theta -\alpha}{2}\bigg)^2\nonumber\\
&=&\frac{4}{{\lambda}_M}+\frac{4}{{\lambda}_M^2}+\frac{4}{{\pi}^2{\lambda}_M^2}\int_0^{\pi}dx({\pi -x})\cos 2x \ln(\sin x)^2+c_2%
\nonumber\\
&=&\frac{4}{{\lambda}_M}+\frac{2}{{\lambda}_M^2}+c_2.
\end{eqnarray}
The specific heat in this phase is therefore given by 
\begin{eqnarray}
\frac{C_{\rm v}}{N^2}=(\frac{2}{{\lambda}_M})^2.\label{lkjo}
\end{eqnarray}

\paragraph{The phase diagram}
From equation (\ref{u}) we can immediately see that  for every value of the parameter $M^2/N$ the model (\ref{gammad4}) undergoes a third order phase transition from strong-coupling phase to weak-coupling phase consistent with the ordinary one-plaquette third order phase transition observed in $2$ dimensional gauge theory on the lattice \cite{gross}. However in this case this transition occurs at the value ${\lambda}_*^u$ of the gauge coupling constant $\lambda$ which becomes smaller as we increase $M^2/N$ until it vanishes for $M^2/N{\longrightarrow}{\infty}$. By definition the double scaling limit in which the fixed ratio $M^2/N$ is taken to infinity 
corresponds to the model (\ref{gammad2}). Indeed the partition function (\ref{gammad2})  is obtained from the partition function  $Z[\gamma,\lambda,M^2]$ given by (\ref{gammad3}) in the limit in which we take $M^2{\longrightarrow}\infty$ first and then $N{\longrightarrow}\infty$. Hence 
we can immediately conclude that the model (\ref{gammad2}) does not undergo the above one-plaquette phase transition since ${\lambda}_*^u{\longrightarrow}0$ when $M^2/N{\longrightarrow}\infty$  and furthermore  this model according to the restriction  (\ref{range5})   exists mostly in the matrix phase. 

Thus the model (\ref{gammad4}) with finite values of the ratio $M^2/N$ behaves as an ordinary $2d$ lattice gauge model with a reduced one-plaquette critical value whereas in the limit $M^2/N{\longrightarrow}\infty$ the $3$rd order one-plaquette phase transition  is completely removed from the model. This limit can be thought of as a way of regularizing the one-plaquette transition in the lattice model.


However for $M^2/N{\longrightarrow}0$ the above computed critical value ${\lambda}_*^u=2$ is not the correct critical value  at which the third order phase transition should occur. 
 The  model (\ref{gammad3}) with the fixed ratio $M^2/N$ such that $M^2/N{\longrightarrow}0$ is a $U(2N)$ one-plaquette model 
which should be described by the partition function (\ref{gammad4-0}) and not by (\ref{gammad4}). By going through the same steps which led to (\ref{u}) we obtain for the model (\ref{gammad4-0}) the critical value ${\lambda}_{M*}=1$ or equivalently 
\begin{eqnarray}
{\lambda}_{*}^l=\frac{1}{1+\frac{M^2}{N}}=1-\frac{M^2}{N}+O(\frac{M^4}{N^2}).
\end{eqnarray}
In above we have used the fact that the partition function (\ref{gammad4-0}) should really be used only for very small $M^2/N$. Clearly for $M^2/N{\longrightarrow}1$ the critical value ${\lambda}_*^l{\longrightarrow}0$ and then it becomes negative  for $M^2/N{>}1$ and hence the above formula should only be valid in the range $M^2/N{\in}[0,1[$. In other words the $3$rd order  one-plaquette phase transition occurs at the values ${\lambda}_*^l$ of $\lambda$ for all $M^2/N{\in}[0,1[$ whereas for $M^2/N{\in}[1,\infty[$ the transition should occur at the values ${\lambda}_*^u$. The value $M^2/N=1$ seems to demarcate a  discontinuity in the phase diagram. Below  $M^2/N=1$ the model (\ref{gammad3}) describes an ordinary $U(2N)$ gauge theory on one plaquette whereas above this  value the model becomes the $U(N)$ gauge theory on two plaquettes given by (\ref{gammad4}) which corresponds to noncommutative gauge theory on the fuzzy sphere. The numbers of degrees of freedom of the model below and above this critical mass ${M^2}/{N}=1$ are different. Above $M^2/N=1$ we have $2N^2$ degrees of freedom whereas below this value we have $4N^2$ degrees of freedom.

Thus the model (\ref{gammad3}) with $M^2/N=1$  is seen to be  the first noncommutative $U(1)$ gauge model on the fuzzy sphere ${ S}^2_N$ which we will obtain as we increase $M^2/N$ from $0$ to $\infty$. Indeed this model has the correct number of degrees of freedom given by $2N^2$ and it can be reduced to (\ref{gammad4}).  The model  with $M^2/N=1$  is also the noncommutative model with the largest ``fuzzy sphere'' phase in the sense of (\ref{range5}) hence  the effects of the ``matrix'' phase are the weakest in this case. As a consequence the estimation of the critical point of the $3$rd order phase transition is  most reliable using  the model (\ref{gammad3}) with $M^2/N=1$. At $M^2/N=1$ the critical value ${\lambda}_*^u$ is given by ${\lambda}_*^u=2/3=0.66$. This computed  critical value leads to the critical value of the coupling constant $\bar{\alpha}$ ( from equation (\ref{mmm}) )
\begin{eqnarray}
\bar{\alpha}_*^4=\frac{64}{{\lambda}_*^u}=96~{\Leftrightarrow}~\bar{\alpha}_*=3.13.\label{3.13}
\end{eqnarray}
This is to be compared with the observed value $\bar{\alpha}_*=3.35{\pm}0.25$ seen in Monte Carlo simulation of the model (\ref{main2})   with the Metropolis algorithm. The agreement is very good. 
Thus it seems that the two noncommutative $U(1)$ models given by (\ref{main2}) ( with large $m^2$ ) and  (\ref{gammad3})  ( with $M^2/N=1$ ) are  quantum mechanically equivalent in  the fuzzy ${S}^2_N$ phase.

\begin{figure}
\begin{center}
\includegraphics[width=8cm,angle=-90]{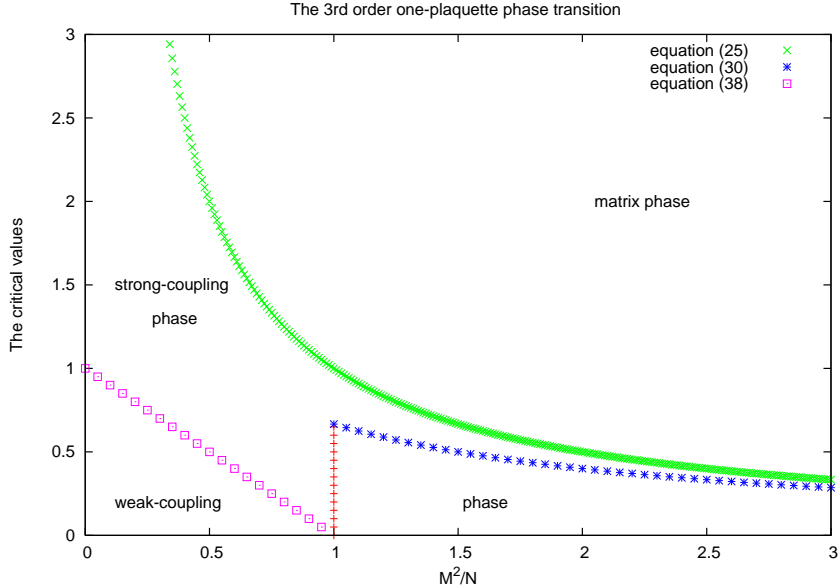}
\caption{The weak-coupling and strong-coupling phases of the noncommutative $U(1)$ gauge theory on the fuzzy sphere are the regions of the weak-coupling and strong-coupling phases with $M^2/N{\geq}1$. The regions  with $M^2/N{<}1$ correspond to an ordinary $U(2N)$ gauge theory. The number of degrees of freedom of the model above the critical mass ${M^2}/{N}=1$ is $2N^2$ whereas below  ${M^2}/{N}=1$ we have $4N^2$ degrees of freedom. The fuzzy sphere phase consists of the weak-coupling and strong-coupling phases with $M^2/N{\geq}1$. In this phase the theory is a $U(N)$ gauge theory on two plaquettes.}
\end{center}\label{phase}
\end{figure}

\section{Conclusion}

We constructed in this article a one-parameter family of noncommutative $U(1)$ gauge theory partition functions $Z_M\equiv Z(\gamma,\lambda,M^2)$ on the fuzzy sphere $S^2_N$  given by equation (\ref{gammad3}). The idempotent $\gamma$ is the chirality operator and $\lambda$ is the gauge coupling constant. In the limit $M^2{\longrightarrow}0$ the model becomes a $U(2N)$ one-plaquette model whereas in the limit  $M^2{\longrightarrow}\infty$ we get the  generalized $U(N)$ one-plaquette model defined by (\ref{gammad2}). In the double scaling limits $M^2,N{\longrightarrow}\infty$ keeping $M^2/N$ fixed ${\geq}1$ we can show that quantum noncommutative $U(1)$ gauge theory on the fuzzy sphere ${ S}^2_N$ is equivalent to a  $U(N)$ nonabelian Yang-Mills theory on a two dimensional lattice with two plaquettes according to (\ref{gammad4}). Thus the model (\ref{gammad3})  ( with $M^2/N=1$ ) will undergo in the fuzzy sphere phase a $3$rd order one-plaquette large $N$ phase transition between weak-coupling and strong-coupling phases around the critical value (\ref{3.13}) which is consistent  with the observed $3$rd order transition point of (\ref{main2}). In the ``matrix'' phase we will have a $U(N)$ gauge theory on a single point.

Another evidence for this equivalence comes from the calculation of the specific heat which seems to agree with the simulation results of (\ref{main2}) both in the weak-coupling and  strong-coupling phases up to the sphere-to-matrix first order transition where the whole spactime ( the sphere ) collapses. 
In particular the specific heat was found to be equal to $1$ in the fuzzy sphere weak-coupling phase of the gauge field which agrees with the observed value $1$. The value $1$ comes precisely because we have two plaquettes which approximate the noncommutative $U(1)$ gauge  field on the fuzzy sphere. In the strong-coupling region deviations between (\ref{lkjo}) and the data coming from the simulation of (\ref{main2}) become significant only near the sphere-to-matrix transition point \cite{ref11,ref1}.



\bibliographystyle{unsrt}

\begin{thebibliography}{99}

\bibitem{thesis}
Badis Ydri,{\it Mod.Phys.Lett.}{A 19} (2004)2205-2213,{\it Nucl.Phys.}{B 690} (2004)230-248.

\bibitem{madore} J.Madore,{\it Class.Quant.Grav.} 9:69-88,1992.
J.Hoppe, MIT PhD thesis (1982). 


\bibitem{ars}
A.Y.Alekseev , A.Recknagel, V.Schomerus, {\it hep-th/0003187} and {\it hep-th/9812193} .

\bibitem{iso}
S.Iso, Y.Kimura, K.Tanaka, K.Wakatsuki, {\it Nucl.Phys.} B604 (2001) 121.


\bibitem{ref11} D.O'Connor, { Badis Ydri} , {\sl Monte Carlo Simulation of NC Gauge Field on the Fuzzy Sphere},  {\it hep-lat/0606013}.


\bibitem{ref1} { Badis Ydri} , {\sl The one-plaquette model limit of NC gauge theory in 2D}, {\it hep-th/0606206}.








\bibitem{ref} P.Castro-Villarreal , R.Delgadillo-Blando , { Badis Ydri}, hep-th/0405201 , {\it Nucl.Phys.B} {\bf 704} (2004) 111-153.





\bibitem{nishimura} T.Azuma,S.Bal,K.Nagao,J.Nishimura,{\it JHEP} 0405 (2004) 005.


\bibitem{steinackers2} H.Steinacker, {\it Nucl.Phys.}{B679},66 (2004). See also the recent preprint : H.Steinacker, R.J. Szabo, {\it hep-th/0701041}. 

\bibitem{gross}
D.J.Gross, E.Witten, {\it Phys.Rev.} D 21 (1980)446-453. S.R.Wadia, EFI preprint  EFI 80/15 ( March 1980 ),
  Phys.\ Lett.\ B {\bf 93} (1980) 403.
 













 








 





\end{thebibliography}

\end{document}